\documentclass[aps,pre,groupedaddress,twocolumn]{revtex4}

\newcommand{\Uc}{\mathcal{U}}

\usepackage{hyperref}
\usepackage{subfigure}
\usepackage{color}
\usepackage{graphicx}
\usepackage{natbib}
\usepackage{doi}

\begin{document}
\bibliographystyle{plainnat}

\expandafter\ifx\csname urlprefix\endcsname\relax\def\urlprefix{URL }\fi

\title{\large
 Hydrodynamics of the Physical Vacuum:
 I.~Scalar quantum sector}

\large
\author{Valeriy I. Sbitnev}
\email{valery.sbitnev@gmail.com}
\address{St. Petersburg B. P. Konstantinov Nuclear Physics Institute, NRC Kurchatov Institute, Gatchina, Leningrad district, 188350, Russia;\\
 Department of Electrical Engineering and Computer Sciences, University of California, Berkeley, Berkeley, CA 94720, USA
}


\date{\today}

\begin{abstract}
  Physical vacuum is a special superfluid medium. Its motion is described by the Navier-Stokes equation
  having two slightly modified terms that relate to internal forces.
  They are the pressure gradient and the dissipation force because of viscosity. 
  The modifications are as follows: 
 (a)~the pressure gradient contains an added term describing the pressure multiplied by the entropy gradient; 
 (b)~time-averaged viscosity is zero, but its variance is not zero.
 Owing to these modifications, the Navier-Stokes equation can be reduced to the Schr\"{o}dinger equation describing 
 behavior of a particle into the vacuum,
 which looks like a superfluid medium populated by enormous amount of virtual particle-antiparticle pairs.


\end{abstract}

\maketitle

\section{Introduction}

 For about 100 years of the study of quantum-mechanical phenomena, the Schr\"odinger equation has proved to be a reliable equation.
 Its solutions give good agreement  with experiments. The solution of this equation is represented through a complex-valued function $\psi({\vec r},\,t)$,
 called the wave function.
 Observable variables, $A$ and others, are expressed through scalar product   $\langle\psi|A|\psi\rangle$,
which represents the statistical expectation of this variable.
 From the birth of the quantum mechanics as a science,  one question arose: what can be the physical meaning of the wave function?
 
 Without going into details of various interpretations of the quantum mechanics~\cite{Hartle2005},
 we note that immediately behind the famous article of Erwin Schr\"odinger published
 in 1926~\cite{Schrodinger1926}, in the same year an article of Erwin Madelung appeared in the German magazine. 
 This article was entitled
 "Quantum theory in hydrodynamical form"~\cite{Madelung1926}. 
 Madelung compares the wave function represented in the polar form $\psi = R\exp\{{\bf i}S/\hbar\}$
 with a peculiar fluid,  which flows according to laws described in this article.
 Due to such a representation of the wave function, the Schr\"odinger equation for the complex-valued function, $\psi$, can be reduced to two equations, each dealing with the real-valued functions $S$ and $R$.
  The first equation is the Hamilton-Jacobi equation loaded by the extra term having a view
\begin{equation}
\label{eq=1}
    Q = -{{\hbar^2}\over{2m}}{{\nabla^{\,2}R}\over{R}}.
\end{equation}
 And the second equation is the continuity equation for the probability density $\rho = R^{\,2}$. Here $R$ is the probability amplitude,
 $\hbar$ is the reduced Planck constant, and $m$ is the mass of a particle.
 It is instructive to note, while the amplitude $R$ relates to the probability to discover a particle within the volume $\Delta V$, the action $S$ describes a mobility of the same particle within this volume.

 Similar computations were made by David Bohm in 1952 and published in the journal Physical Review~\cite{Bohm1952a, Bohm1952b}. Immediately following these publications, David Bohm and Jean-Pierre Vigier published an article concerning a causal interpretation of quantum mechanics, in which a set of fields is equivalent in many ways to a conserved fluid~\cite{BohmVigier1954}.
 Interest to searching analogies between quantum mechanical phenomena and hydrodynamical phenomena
 always stays heightened~\cite{BuzeaEtAl2012, DamazioSiqueira2013, Harvey1966, Holland2005, JungelMili2012, Wyatt2005}.

Appropriate to note the approaches to quantum mechanics which begin with the primary principles of statistical mechanics and nonequilibrium thermodynamics~\cite{FritscheHaugk2009, Grossing2010, Nelson1966}.
The problem here is to describe correctly behavior of matter in small space-time volumes  populated by enormous amount of sub-particles. 
The latter behave like  the Brownian particles~\cite{Nelson1967, Nelson1985}.

 After Bohm's publications~\cite{Bohm1952a, Bohm1952b}, the term~(\ref{eq=1}) is named the quantum potential and the computations given in these articles are represented further as the Bohmian mechanics~\cite{BensenyEtAl2014, LicataFiscaletti2014b}.
 Bohm's version of quantum mechanics was practically a mathematical interpretation of the de Broglie pilot-wave theory.
 According to this theory, there is a wave accompanying the particle from its birth on a source until it will be registered on a detector~\cite{deBroglie1987}.
 The particle is represented as a singularity moving along the optimal trajectory, formed due to the interference of
 the pilot-wave accompanying particle.

Surprisingly, the de Broglie theory being proposed for explanation of quantum mechanical phenomena, finds its application in an unexpected realm: at studying the motion of droplets on the oil surface, scientists have observed interference of surface waves induced by the bouncing droplets~\cite{CouderForte2006, EddiEtAll2011, ProtiereEtAll2006}. 
It turns, these waves guide the droplets along Bohmian-like paths,
so that the observed interference fringes are similar to those registered for electrons passing through a screen with two slits. 
An excellent overview of the movement of droplets on the oil surface, where subcritical Faraday waves are supported, is given in Bush's article~\cite{Bush2015a}. There is given comparison of the described phenomena with similar phenomena observed in quantum mechanics.

 Let us imagine that the subcritical Faraday waves induced on the fluid surface  simulate, in some way, the zero-point fluctuations of the physical vacuum. In other words, the physical vacuum manifests itself as a special fluid - the superfluid one~\cite{Donnelly1991}, and~\cite{Volovik2003}. For that reason we may apply the Navier-Stokes equation for description of the state of such a fluid~\cite{Sbitnev2014a}, and~\cite{Sbitnev2015b}, where the viscosity in the first approximation vanishes.

Two equations which have a crucial significance here are the Navier-Stokes equation~\cite{Sbitnev2015b} 
\begin{equation}
 \rho_{_{M}}\biggl(
 {{\partial {\vec {\mathit v}}}\over{\partial\,t}}
 + ({\vec {\mathit v}}\cdot\nabla){\vec {\mathit v}}
       \biggr) 
  = {{{\vec{\mathit F}}}\over{\Delta V}}
 - \rho_{_{M}}\nabla \biggl( {{P}\over{\rho_{_{M}}}} \biggr)
  + \mu(t)\,\nabla^{\,2}{\vec {\mathit v}}.
\label{eq=2}
\end{equation}
 and  the continuity equation
\begin{equation}
 {{\partial\,\rho_{_{M}}}\over{\partial\,t}} +(\nabla\cdot{\vec{\mathit v}})\rho_{_{M}} = 0.
\label{eq=3}
\end{equation}
 Here $\rho_{_{M}}=M/\Delta V$ is a mass density of the fluid in the volume $\Delta V$,
 ${\vec{\mathit v}}$ is the flow velocity, and ${\vec{\mathit F}}/\Delta V$ is an external force per the volume $\Delta V$.
There are internal forces that are represented by pressure gradients within the fluid and dissipative forces arising due to the fluid viscosity. They are represented by the last two terms in Eq.~(\ref{eq=2}). The following modifications of these terms are important in this article: (a) the dynamic viscosity, $\mu$, fluctuates in time about its zero value; (b) the pressure gradient is subjected to a slight modification, namely 
 $\nabla P \rightarrow \rho_{_{M}}\nabla(P/\rho_{_{M}})=\nabla P - P \nabla\ln(\rho_{_{M}})$.
 This modification will be important for us when we shall begin to derive the Schr\"odinger equation.

 The article is organized as follows. Sect.~\ref{sec2} deals with consideration of the Navier-Stokes equation in the light of derivation of the Schrodinger equation. A main point here lies in finding the quantum potential~(\ref{eq=1}).
Sect.~\ref{sec3} discusses the Bohmian trajectories and the uncertainly principle related to them.
 In Sect.~\ref{sec4}, the concluding section, we consider different interpretations of quantum mechanics in the light of these findings.
As for the vorticity dynamics in the fluid with the fluctuating viscosity, this item is considered in the following article.

\section{\label{sec2}Derivation of the Schr\"odinger equation}

 First let us rewrite the mass density $\rho_{_{M}}$ according to the following presentation
\begin{equation}
 \rho_{_{M}} = {{M}\over{\Delta V}}={{mN}\over{\Delta V}} = m\rho.
\label{eq=4}
\end{equation}
Here the mass $M$ of the fluid that fills the volume $\Delta V$ is equal to
the product of the number,  $N$, of identical sub-particles populating this volume, 
by the mass $m$ of these sub-particles~\cite{JackiwEtAl2004}.
 Consequently, the mass density $\rho_{_{M}}$ is  the product of
 the mass $m$ by the density of the sub-particles $\rho=N/\Delta V$.
 
 We believe that the sub-particles perform Brownian motions in the fluid medium.
 The diffusion coefficient of this motion has the following view
\begin{equation}
\label{eq=5}
   D = {{\hbar}\over{2m}}.
\end{equation}
The Brownian motion of the sub-particles represents an important essence in Nelson's work~\cite{Nelson1966}. It determines states of aether, wherein quantum events happen. For that description Nelson attracts methods of statistical mechanics: the white noise through the Wiener process initiates uncertain directions of the motion.
 In our case, this fluid medium is the physical vacuum, being populated by enormous numbers of virtual particle-antiparticle pairs. Colliding particles obey to the uncertainty principle: momenta of the particles after the collision acquire uncertain values.

\subsection{\label{subsec2A}Where does the quantum potential come from?}

 We begin with the modified pressure gradient
\begin{equation}
\label{eq=6}
  \rho_{_{M}}\nabla\biggl(
                   {{P}\over{\rho_{_{M}}}}
                   \biggr)
=   \rho\nabla\biggl(
                   {{P}\over{\rho}}
                   \biggr)
 = \nabla P - P\nabla\ln(\rho_{_{}}).
\end{equation}
 The first term, the pressure gradient $\nabla P$, is represented in the usual Navier-Stokes equation~\cite{LandauLifshitz1987}, and~\cite{KunduCohen2002}. 
 Whereas, the second term, $P\nabla\ln(\rho_{_{}})$, is an extra term describing change in the logarithm of the density distribution $\rho$ on increment of length multiplied by $P$. 
 It means that change of pressure is induced by change of entropy per length, or else
 by change of information flow~\cite{FiscalettiLicata2014},~\cite{LicataFiscaletti2014a}~\cite{Sbitnev2009} per length. 

 First, let us suppose that the pressure $P$ is the sum of two pressures $P_1$ and $P_2$. As for the pressure $P_1$ we begin with the Fick's law~\cite{FritscheHaugk2009},~\cite{Grossing2010}. 
 The law says that the diffusion flux, $\vec{\mathit J}$, is proportional to the negative value of the density gradient, ${\vec{\mathit J}} =-D\nabla\rho_{_{M}} $. Next, since the term ${D}\nabla{\vec{\mathit J}}$
has dimension of the pressure, we define:
\begin{equation}
\label{eq=7}
  P_1 = {D}\nabla{{\vec{\mathit J}}}
 = -{{\hbar^{2}}\over{4m^2}}\nabla^{2}\rho_{_{M}}.
\end{equation}
 Here we  replace the diffusion coefficient $D$ by its expression shown in Eq.~(\ref{eq=5}).
 Observe that the kinetic energy of the diffusion flux is $(m/2)({\vec{\mathit J}}/\rho_{_{M}})^{2}$.
 It means that there exists one more pressure
 as the average momentum transfer per unit area per unit time:
\begin{equation}
\label{eq=8}
  P_2 = {{\rho_{_{M}}}\over{2}}\biggl(
                                      {{{\vec{\mathit J}}}\over{\rho_{_{M}}}}
                               \biggr)^2
 = {{\hbar^2}\over{8m^2}}{{(\nabla\rho_{_{M}})^2}\over{\rho_{_{M}}}}.
\end{equation}
 Now we can see that sum of the two pressures, $P_1 + P_2$, divided by $\rho$ (we remark that $\rho_{_{M}} = m\rho$, see Eq.~(\ref{eq=4}))  reduces to the quantum potential
\begin{equation}
\label{eq=9}
  Q =  {{P_{2}+P_{1}}\over{\rho}}
     = {{\,\hbar^2}\over{8m}}\biggl(
                                   {{\nabla\rho}\over{\rho}} 
                                   \biggr)^2
 - {{\,\hbar^2}\over{4m}}{{\nabla^2\rho}\over{\rho}} .
\end{equation}
            
 Now the Navier-Stokes equation~(\ref{eq=2}) divided by $\rho$ takes a view
\begin{equation}
 m\biggl(
 {{\partial {\vec {\mathit v}}}\over{\partial\,t}}
 + ({\vec {\mathit v}}\cdot\nabla){\vec {\mathit v}}
       \biggr) 
  =  {{{\vec{\mathit F}}}\over{N}}
   \;-\; \nabla Q
 \; +\; \nu(t)\,\nabla^{\,2}m{\vec {\mathit v}}.
\label{eq=10}
\end{equation}
 Here ${\vec{\mathit F}}/N$ is the external force per the sub-particle. 
 The kinetic viscosity $\nu(t)=\mu(t)/\rho_{_{M}}$ has the dimension [length$^2$/time] and fluctuates  about zero. Let its amplitude be equal to the diffusion coefficient~(\ref{eq=5}).

\subsection{\label{subsec2B}Irrotational and solenoidal vector fields}

 The fundamental theorem of vector calculus states that any vector field can be expressed through sum of irrotational and solenoidal fields.
The current velocity ${\vec{\mathit v}}$ can be represented consisting of two components --  irrotational and solenoidal~\cite{KunduCohen2002}:
\begin{equation}
\label{eq=11}
  {\vec{\mathit v}} = {\vec{\mathit v}}_{_{S}} + {\vec{\mathit v}}_{_{R}},
\end{equation}
 where subscripts $S$ and $R$ point to existence of scalar 
  and vector (rotational) potentials  underlying emergence of these  velocities.
 These velocities relate to vortex-free
 and vortex motions of the fluid medium, respectively. Scalar and vector fields underlie of manifestation of these two types of the velocities.
 They satisfy the following equations
\begin{equation}
\label{eq=12}
 \left\{
    \matrix{
           (\nabla\cdot{\vec{\mathit v}}_{_{S}}) \ne 0, & [\nabla\times{\vec{\mathit v}}_{_{S}}]=0, \cr
           (\nabla\cdot{\vec{\mathit v}}_{_{R}})  =  0, &\, [\nabla\times{\vec{\mathit v}}_{_{R}}]={\vec\omega}. \cr
           }
 \right.
\end{equation}
 Also we shall write down the term $({\vec{\mathit v}}\cdot\nabla){\vec{\mathit v}}$ in Eq. (\ref{eq=10})  in detail
\begin{equation}
\label{eq=13}
  ({\vec{\mathit v}}\cdot\nabla){\vec{\mathit v}} =
 \nabla {\mathit v}^2 /2   +   [{\vec\omega}\times{\vec{\mathit v}}].
\end{equation}
 Here ${\vec\omega} = [\nabla\times{\vec{\mathit v}}]$ is the vorticity.
 
 Let us now designate the scalar field by a scalar function $S$, and the name is the action.
 Then the irrotational velocity ${\vec{\mathit v}}_{_{S}}$ of the sub-particle is defined as $\nabla S/m$.
 Next, the momentum and the kinetic energy of the sub-particle  have the following representation
\begin{equation}
\label{eq=14}
 \left\{\,
    \matrix{
           {\vec{\mathit p}} = m{\vec{\mathit v}} = \nabla S + m{\vec{\mathit v}}_{_{R}}, \cr\cr
           {\displaystyle
            m{{{\mathit v}^2}\over{2}} = 
           {{1}\over{2m}}(\nabla S)^2 + m {{{\mathit v}_{_{R}}^2}\over{2}}}. \cr
           }
 \right.
\end{equation}
 As seen in the bottom line, the kinetic energy of the fluid motion is equal to sum of the kinetic energies of irrotational and solenoidal motions.

 Now taking into account the expression~(\ref{eq=13})
 we may rewrite the Navier-Stokes equation~(\ref{eq=10}) in the more detailed form
\begin{eqnarray}
\nonumber
&&
    {{\partial}\over{\partial\,t}}(\nabla S + m{\vec{\mathit v}}_{_{R}})
\\ \nonumber
   &+& 
  \biggl\{ {{1}\over{2m}}\nabla\Bigl(    
                      (\nabla S)^2 + m^2{\mathit v}_{_{R}}^2
                      \Bigr)
    +   m\bigl[{\vec\omega}\times
                {\vec{\mathit v}}_{_{R}}
        \bigr]
 \biggr\}  
\\ \nonumber
\\
   &=& - \nabla U - \nabla Q
    + 
          \nu(t)\nabla^2(\nabla S + m{\vec{\mathit v}}_{_{R}}).
\label{eq=15}
\end{eqnarray}
 Here we take into account that the external force in the Navier-Stokes equation~(\ref{eq=10}) is conservative,
 i.e.,  ${\vec F}/N = -\nabla U$, where $U$ is the potential energy relating to the single sub-particle.
 The terms $\nabla U$ and $\nabla Q$ are gradients of the potential energy and of the quantum potential, respectively.

 The third term in the second line describes the viscosity of the medium. 
 As was said above, time-averaged the viscosity vanishes.
 Its variance is not zero.
 Because of this term the above equation is akin to the Langevin equation
with $\delta$-correlation function $\nu(t)$.

 Now we rewrite the above equation 
  by regrouping its terms also
\begin{widetext} 
\begin{eqnarray}
 \nabla\biggl(
  \underbrace {
  {{\partial~}\over{\partial\,t}}S + {{1}\over{2m}}(\nabla S)^2 + {{m}\over{2}}{\mathit v}_{_{R}}^2
              + U + Q - \nu(t)\nabla^{2}S}_{f_1({\vec r},t)}
       \biggr)
 =  \underbrace {
 -m{{\partial~}\over{\partial\,t}}{\vec{\mathit v}}_{_{R}}
   - m [\,{\vec\omega}\times{\vec{\mathit v}}_{_{}}\,]
    +\nu(t)m\nabla^2 {\vec{\mathit v}}_{_{R}}}_{{\vec f}_2({\vec r},t)}.
\label{eq=16}
\end{eqnarray}
\end{widetext}
Here the curly brackets single out two qualitatively different functions, scalar-valued fuction $f_1({\vec r}, t)$ and vector-valued function 
${\vec f}_2({\vec r}, t)$. Depending on application of the $\nabla$-operator, either as inner product of this operator or the cross product, we may extract either the left side or the right side of this expression.
This equation is  almost similar to that given in~\cite{FritscheHaugk2009}. 
The exceptions are introduction of the fluctuating viscosity $\nu(t)$
  ({
   here we have absolutely different insights}).
The vector terms is marked by the curly bracket ${\vec f}_{2}({\vec r}, t)$. 
The latter have crucial signification for describing the vortices.
The vortex dynamics will be described in the next article.

 Let us multiply Eq.~(\ref{eq=16})  by the  curl.
We find:
\begin{equation}
\label{eq=17}
 \nabla\times \nabla f_1({\vec r}, t)  = 0.
\end{equation}
It means that the function $ f_1({\vec r}, t)$  is any scalar function of ${\vec r}$ and $t$,
   $ f_1({\vec r}, t) =  C$, and  $C$ is an arbitrary integration constant.
 As a result, we come to the following modified Hamilton-Jacobi equation 
\begin{equation}
\label{eq=18}
    {{\partial~}\over{\partial\,t}}S + {{1}\over{2m}}(\nabla S)^2 + {{m}\over{2}}{\mathit v}_{_{R}}^2
      + U 
      + Q 
      - \underbrace{\nu(t)\nabla^{2}S}_{(a)}
      = C.
\end{equation}
 The modification of this equation is due to the presence of the quantum potential $Q$,
 that is shown in Eq.~(\ref{eq=9}).
 The second equation, in addition to the modified Hamilton-Jacobi equation, is the continuity equation
\begin{eqnarray}
\nonumber
     {{\partial\,\rho_{_{}}}\over{\partial\,t}} +(\nabla{\vec{\mathit v}})\rho_{_{}}
 &=& {{\partial\,\rho_{_{}}}\over{\partial\,t}} +  ({\vec{\mathit v}}\nabla)\rho_{_{}}
    + \rho_{_{}}(\nabla{\vec{\mathit v}}) \\
&=&  {{d\,\rho_{_{}}}\over{d\,t}}
     + \rho_{_{}}(\nabla{\vec{\mathit v}})
 = 0
\label{eq=19}
\end{eqnarray}
rewritten from Eq.~(\ref{eq=3}) in accordance with~(\ref{eq=4}).
 Eqs.~(\ref{eq=18}) and~(\ref{eq=19}) are seen to be crucial equations in the Bohmian mechanics~\cite{Bohm1952a, Bohm1952b}, and~\cite{BensenyEtAl2014}.

The both real-valued Eqs.~(\ref{eq=18}) and~(\ref{eq=19}) give sufficient conditions  for transition to the complex-valued Schr\"odinger equation. However an extra-term, engulfed by the curly bracket (a) in the first equation, breaks such a transition. Nevertheless  by using the continuity equation~(\ref{eq=19})  we may rewrite the term $\nabla^{2}S$ in the following convenient form
\begin{equation}
  \nabla^{2}S = m(\nabla{\vec{\mathit v}})
   = - m{{d\,}\over{d\,t}}\ln(\rho)
   = - m f(\rho).
\label{eq=20}
\end{equation}
 Now we may rewrite Eq.~(\ref{eq=18}):
\begin{equation}
\label{eq=21}
    {{\partial~}\over{\partial\,t}}S + {{1}\over{2m}}(\nabla S)^2 + {{m}\over{2}}{\mathit v}_{_{R}}^2
      + U 
      + Q 
      + \underbrace{\nu(t)m f(\rho)}_{(a)}
      = C.
\end{equation}
 The function $f$ can be expanded into Tailor series as a polynomial in $|\Psi|^2 = \rho$
 with real coefficients~\cite{AbidEtAl2003}.

Observe that the similar quantum Hamilton-Jacobi equation stems from 
the Gross-Pitaevskii (GP) equation, when the Madelung transformation
\begin{equation}
\label{eq=22}
  \Psi = \sqrt{\,\rho\,}\exp\{{\bf i}S/\hbar\}
\end{equation}
is applied to the GP equation~\cite{AbidEtAl2003, RobertsBerloff2001}.
In these works the GP equation is used for description of the Bose-Einstein condensate in superfluid $^4$He.
 In our case we have an inverse problem - we need to go from
  the equations~(\ref{eq=19}) and~(\ref{eq=21}) to the GP-like equation
\begin{eqnarray}
\nonumber
  {\bf i}\hbar\,{{\partial\Psi}\over{\partial\,t}} &=&
  {{1}\over{2m}}(-{\bf i}\hbar\nabla + m{\vec{\mathit v}}_{_{R}})^2\Psi
      + \underbrace{\nu(t)m f(\rho)}_{(a)}\Psi  \\
      &+& U({\vec r})\Psi  
       - C\Psi
\label{eq=23}
\end{eqnarray}
  which is the nonlinear  Schr\"odinger equation, because of existence of the term marked by (a).
 The kinetic momentum operator $(-{\bf i}\hbar\nabla + m{\vec{\mathit v}}_{_{R}})$  contains the term $m{\vec{\mathit v}}_{_{R}}$ describing
 a contribution of the vortex motion. This term is analogous to the vector potential~\cite{BerryEtAl1980, Martins2012} multiplied by the ratio of the charge to the light speed,
 which appears in quantum electrodynamics. 

In the quantum realm, the kinetic viscosity coefficient, $\nu(t)$, can be rewritten in a more detailed form
\begin{equation}
  \nu(t) = {{\hbar}\over{2m}}g(t).
\label{eq=24}
\end{equation}
Here the coefficient $\hbar/2m$ is the same as shown in Eq.~(\ref{eq=5}) and $g(t)$ is a dimensionless function compiled from all possible oscillating modes represented by creation and annihilation operators of the virtual particles.
It is seen that we can make the following replacement of the wave function:
\begin{equation}
   \Psi = \psi\cdot \exp\Bigg\{
                                        -{{\bf i}\over{2}}\int\limits_{0}^{t} g(\tau)f(\rho) d\tau
                                \Biggr\}.
\label{eq=25}
\end{equation}
In this case the term marked by the curly bracket (a) in~(\ref{eq=23}) can be deleted.
Inasmuch as $\nu(t)$ is zero in average on the time, then the integral under the exponent tends to zero for $t$ large enough. Therefore we may ignore this term in the case of laboratory experiments on the Earth.
({
It is interesting to note that manifestation of the viscosity of the physical vacuum on the cosmic scale can give explanation of phenomena of the "Pioneer anomaly"~\cite{Alemanov2015} and the tired light through the quantum Hubble's law~\cite{Alemanov2014}}).

\section{\label{sec3}Bohmian trajectories and the uncertainty principle}

  The Schr\"{o}dinger wave equation can be resolved by heuristic writting of a solution by using
 the Huygens' principle, which mathematically looks as~\cite{Makri1991}:
\begin{equation}
  \Psi({\vec r},t)= \int  K({\vec r},{\vec\xi};t)\Psi({\vec\xi},0)d{\vec\xi}.
\label{eq=26}
\end{equation}
 The propagator $ K({\vec r},{\vec\xi};t)$  contains information on a physical space loaded by physical equipments, such as sources, detectors, collimators, gratings, etc. 
 This physical scene is described by the potential  $U({\vec r})$ 
  and by the boundary conditions superposed on it.
 So, a solution of the Schr\"odinger equation can be achieved by applying
  the Feynman path integral technique~\cite{FeynmanHibbs1965, Derbes1996}.

 By applying this technique for describing the wave propagation through a grating consisting of $N$ slits
 we get the solution~\cite{Sbitnev2012}
\begin{widetext} 
\begin{equation}
  |\Psi(x,z)\rangle =
 {{1}\over{N\sqrt{\displaystyle 1 + {\bf i} {{\lambda z}\over{2\pi b^{2}}} }}} 
 \cdot
   \sum\limits_{n=0}^{N-1}
  \exp
  \left\{
   \matrix{
   - {{\displaystyle \Biggl(x - \Biggl(n - {{N-1}\over{2}}
                                \Biggr)d
                     \Biggr)^{2}}\over{\displaystyle 2b^2\Biggr(
                                                                 1 + {\bf i} {{\lambda z}\over{2\pi b^{2}}}
                                                         \Biggl)}}
          }
  \right\}.
\label{eq=27}
\end{equation}
\end{widetext}
 Here $\lambda$ is the de Broglie wavelength, $b$ is the slit width,
 $d$ is the distance between  slits, and $n$ is sequence number of the slit. 
\begin{figure*}[htb!]
 \centering
  \begin{picture}(200,290)(85,-300)
      \includegraphics[scale=0.5, angle=270]{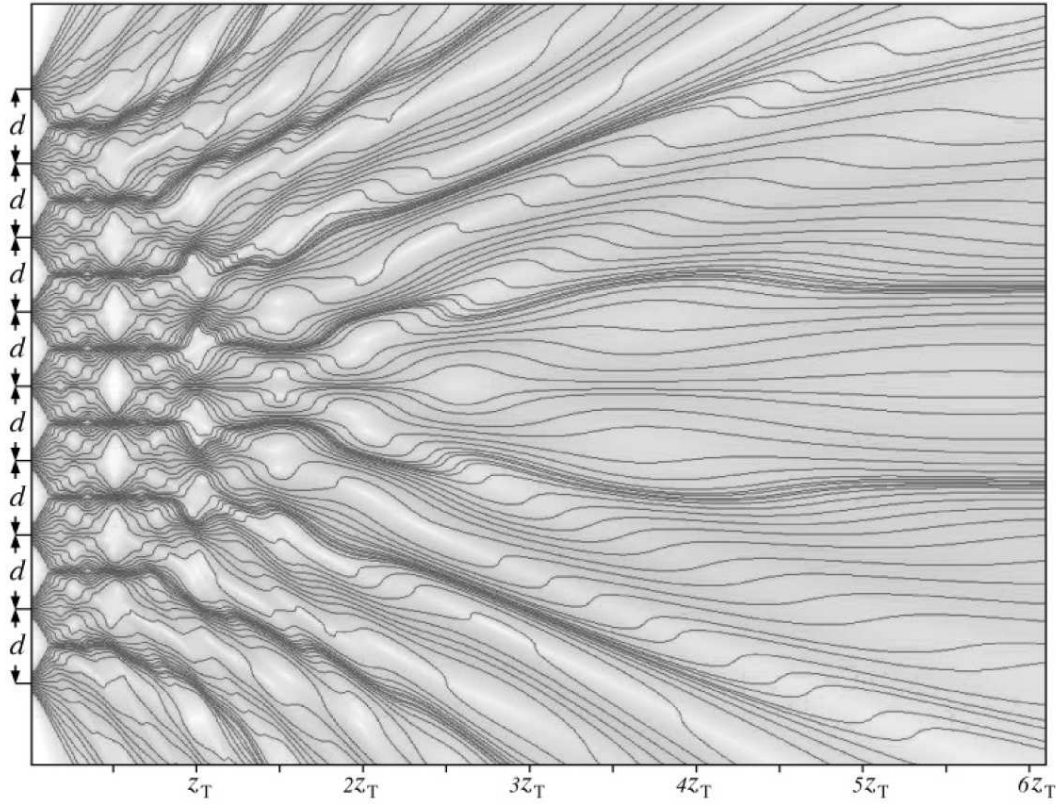}
  \end{picture}
  \caption{
  Interference pattern of the coherent flow of fullerene molecules with de Broglie wavelength $\lambda=5$~pm within
  a zone $z\le 6z_{_{\bf T}}$ from the grating containing $9$ slits.  The scale along the axis $z$ is given in the Talbot lengths, $z_{_{\rm T}} = 2d^{2}/\lambda = 0.025$~m, where  $d = 250$~nm.
 Black curves against the grey background represent the Bohmian trajectories.
   }
  \label{fig=1}
\end{figure*}
 The density distribution function is 
 a scalar product of the wave function $|\Psi(x,z)\rangle$:
\begin{equation}
\label{eq=28}
 p(x,z) = \langle\Psi(x,z)|\Psi(x,z)\rangle.
\end{equation}
 It is shown in gray scale in Fig.~\ref{fig=1}. These calculations show interference of the coherent flow of fullerene molecules~\cite{Sbitnev2013a} from the grating that contains $N=9$ slits. 
The fullerene molecules have the de Broglie wavelength $\lambda = 5$~pm what corresponds to the current velocity of the molecules about 100~m/s~\cite{JuffmannEtAl2009}.
 Parameters of the grating are as follows:  wide of the slit $b=5\cdot10^{3}\lambda=25$~nm, distance between the slits $d=5\cdot10^{4}=250$~nm.

 Black curves drawn against the grey background in Fig.~\ref{fig=1} are 
 the Bohmian trajectories~\cite{Bohm1952a, Bohm1952b}. 
 Their calculations can be found, for instance, in~\cite{BensenyEtAl2014}.  Here we shall calculate these trajectories by adopting the velocity operator that have the form ${\hat{\mathit v}}=-{\bf i}{\hbar/m}\nabla$. For the calculation we use the wave function $|\Psi\rangle$ represented in the polar form~(\ref{eq=22}) with the amplitude density $R$ instead of the probability density $\rho=R^{2}$. The computations give:
\begin{equation}
 {\vec{\Uc}}  = {{1}\over{R^{2}}}
 \langle\Psi\Biggl| -{\bf i}{{\hbar}\over{m}}\nabla\Biggr|\Psi\rangle
 = {{1}\over{m}}\nabla S - {\bf i}{{\hbar}\over{m}}{{\nabla R}\over{R}}.
\label{eq=29}
\end{equation}
 Hereinafter $R^{2} = \langle\Psi|\Psi\rangle$.

 Real part in this expression is the current velocity of the particle
\begin{equation}
  {\vec{\mathit v}} = {\rm Re}\; {\vec{\Uc}} = {{1}\over{m}}\nabla S.
\label{eq=30}
\end{equation}
 Whereas imaginary part 
\begin{equation}
 {\vec{\mathit u}} = -{\rm Im}\; {\vec{\Uc}} = {{\hbar}\over{m}}{{\nabla R}\over{R}}.
\label{eq=31}
\end{equation}
represents the osmotic velocity~\cite{Nelson1966}, which has a deep relation with the quantum potential~(\ref{eq=1}).

 Position of the particle in each current time starting from any slit up to the detector can be calculated by the following increment formula
\begin{equation}
\label{eq=32}
  {\vec r}(\,t+{\delta t}\,) = {\vec r}(\,t\,) +  {\vec{\mathit v}}(\,t\,) {{\delta t}}.
\end{equation}
 Here $t$ is the time that starts at $t=0$ on a source within the slit and ${\delta t}$ is an arbitrarily small increment of the time.
 Some Bohmian trajectories calculated by this method are shown in Fig.~\ref{fig=1} as black wavy curves.

\subsection{\label{subsec3A}The uncertainty principle}

As said, the calculations of~(\ref{eq=30}) and~(\ref{eq=32})
 if executed simultaneously are incompatible with each other because of the uncertainty principle. 
For that reason we need also to evaluate variance of the above computations~\cite{Sbitnev2013a}.
Such an evaluation~is
\begin{eqnarray}
\nonumber
\hspace{-34pt}
 &&{\rm Var}(\vec{\Uc})= {{1}\over{R^{2}}}
 \langle\Psi\Biggl|\Biggl( -{\bf i}{{\hbar}\over{m}}\nabla -\vec{\Uc} \Biggr)^2\Biggr|\Psi\rangle
\\  \nonumber 
\hspace{-34pt} &&
\\
\hspace{-34pt}
&& =  {{1}\over{R^{2}}}
 \langle\Psi\Biggl|
 {\bf i}{{\hbar}\over{m}}\nabla{\vec{\Uc}}
 + \underbrace{
       {\bf i}{{\hbar}\over{m}}{\vec{\Uc}}\nabla
        + {\Uc}^{\,2}
                        }_{(\rm a)}
  - {{\hbar^2}\over{m^2}}\Delta
 \Biggr|\Psi\rangle.
\label{eq=33}
\end{eqnarray}
 The terms over curly bracket (a) kill each other since the operator ${\bf i}(\hbar/m)\nabla$ reproduces $-\vec{\Uc}$, see Eq.~(\ref{eq=29}).
 It is reasonable in the perspective to multiply the above expression by $m/2$:
\begin{equation}
 {{m}\over{2}}{\rm Var}(\vec{\Uc})=
-  {{1}\over{R^{2}}} \langle\Psi\Biggl|
   {{\hbar^2}\over{2m}}\Delta
     \Biggr|\Psi\rangle
 +       {\bf i}{{\hbar}\over{2}}\nabla{\vec{\Uc}}
\label{eq=34}
\end{equation}
So, this expression has a dimensionality of energy.
After a number of computations we get the following result~\cite{Sbitnev2013a}:
\begin{eqnarray}
\hspace{-12pt}
 {{m}\over{2}} {\rm Var}(\vec{\Uc}) 
 ={{m}\over{2}}\Biggr(
 {{1}\over{m}}   \nabla S  -   {\bf i}{{\hbar}\over{m}} {{\nabla R}\over{R}}
                         \Biggl)^{\!2}
                          = {{m}\over{2}}\,{\Uc}^{\,2}.
\label{eq=35}
\end{eqnarray}
 One can see that the term $(m/2){\rm Var}({\vec{\Uc}})$ contains both the kinetic energy of the particle, $(m^2/2)(\nabla S)^2$,
 and the energy of the osmotic motion, $(\hbar^2/2m)(\nabla R/R)^2$.
The generalization of the kinetic energy by expanding it onto the imaginary area belongs the complexified Lagrangian mechanics~\cite{Sbitnev2009}.

 Further we shall consider only real part of the expression over curly bracket~(b):
\begin{equation}
 {{m}\over{2}} {\rm Re\; Var}(\vec{\Uc}) 
 =  {{1}\over{2m}} (\nabla S)^2
 -  {{\hbar}\over{2}}\,\omega_{_{Q}} \ge 0.
\label{eq=36}
\end{equation}
 The first term here is the kinetic energy, $E$, of the particle. And the term
\begin{equation}
 \omega_{_{Q}} = {{\hbar}\over{m}}\Biggr[{{\nabla R}\over{R}}\Biggl]^{\!2},
\label{eq=37}
\end{equation}
stemming from the quantum potential is a frequency of perturbation of the superfluid medium induced by the particle moving through it.
 Let two particles have nearest trajectories.
 For the first particle we have $E_{1} - \hbar\omega_{_{Q,1}}$
 and for the second particle  $E_{2} - \hbar\omega_{_{Q,2}}$. Subtracting one from another we have
\begin{equation}
  {\delta E} - {{\hbar}\over{2}}\,{\delta\omega_{_{Q}}} \ge 0.
\label{eq=38}
\end{equation}
 Two particles moving along the trajectories placed near each other induce perturbations of the kinetic energies during the time ${\delta t}={1/\delta\omega_{_{Q}}}$.
 So that
\begin{equation}
  {\delta E}\,{\delta t} \ge {{\hbar}\over{2}}.
\label{eq=39}
\end{equation}

 Let us return now to Eq.~(\ref{eq=32}) and rewrite it in the following view
\begin{equation}
 {\delta {\vec r}(t)} = {\vec\mathit v}_{1}(t) {\delta t}
 \ge  {\vec\mathit v}_{1}(t) \hbar/2{\delta E}.
\label{eq=40}
\end{equation}
 The initial Bohmian trajectory is marked here by subscript 1.
 Here
 ${\delta E} = m({\mathit v}_{2}^{2} - {\mathit v}_{1}^{2})/2
 \approx m {\vec\mathit v}_{1}{\delta{\vec\mathit v}}$,
 and we  calculated ${\mathit v}_{2}^{2} =
 ({\vec\mathit v}_{1} + {\delta\vec\mathit v})^2
\approx {\mathit v}_{1}^{2} + 2 {\vec\mathit v}_{1}{\delta{\vec\mathit v}}$.
 Substituting computations of ${\delta E}\approx m {\vec\mathit v}_{1}{\delta{\vec\mathit v}}$ 
 into Eq.~(\ref{eq=40}) and taking into account that $m{\delta{\vec\mathit v}}={\delta{\vec\mathit p}}$ we obtain finally
\begin{equation}
  {\delta{\vec\mathit p}}\,{\delta{\vec\mathit r}} \ge {{\hbar}\over{2}.}
\label{eq=41}
\end{equation}

 The uncertainty principle is the fundamental principle of measurements. In this respect the Bohmian trajectories are really "hidden entities" with the point of view of this principle. Attempt to measure attributes of the particle moving along the Bohmian trajectory faces with the uncertainty of position of the particle on the trajectory and its velocity along.  A very small quantity of molecules or atoms in an experimental device (dirt particles) can represent "sources of unpredictable measurements". What leads to deterioration of the visibility of the interference fringes~\cite{JuffmannEtAl2009, Sbitnev2013a}.

\section{\label{sec4}Conclusion}

The Navier-Stokes equation demonstrates a rich set of solutions describing motion of the viscous fluid.
For describing the motions of such a medium, in addition to the force applied from the outside, 
two internal forces should also be taken into account. 
The first force is due to viscosity of the medium because of which dissipation of the stored energy into heat occurs.
The second force comes from the pressure, arising in such a deformable medium.

It turns out that the Navier-Stokes equation can also describe motion of a special superfluid medium - the physical vacuum.
That is, there is the one-to-one mapping the field of the irrotational velocities together with the density distribution of particles having these velocities
onto the complex-valued function, called the wave function.
For that, the above mentioned forces need to be slightly modified.
As a result, we arrive to the Schr\"{o}dinger equation describing evolution of the wave function.

The first force is a dissipative force conditioned by viscosity of the fluid medium. The modification of this term is that we admit that time-averaged the viscosity vanishes, but its variance is not zero. Factually, the viscosity changes its own sign with time. It means that there is an energy exchange within this medium.
 For that reason the fluid medium represents itself as the superfluid vacuum consisting of enormous amount of virtual particle-antiparticle pairs. Really, the superfluid vacuum is represented as the Bose-Einstein condensate.

Modification of the second force should be such that a final product would be the quantum potential. 
For that reason the pressure gradient is supplemented by a term representing the pressure multiplied by gradient of the quantum entropy (gradient of the logarithm from the density distribution of virtual particle-antiparticle pairs).
This latter term, that is responsible for the quantum potential in quantum realm, is obviously absent at description of the classical fluid. 

Inasmuch as the quantum entropy represents the measure of ordering of the virtual pairs in this special medium, 
its gradient describes a flow of the ordered distribution of these pairs to a region where the ordering is in deficit.
It describes an osmotic pressure arising in the physical vacuum.
In the whole, the quantum potential is represented as the ratio of the pressure to the density distribution.
So that, information about a particle state transmitted on enormous distances is achieved due to the osmotic pressure arising in the superfluid medium. Really, it confirms de Broglie's idea~\cite{deBroglie1987} about the wave function as a pilot wave guiding the particle along an optimal path.
Formation of the pilot wave is achieved due to the constructive and destructive interference waves induced by the particle~\cite{FeynmanHibbs1965} passing through the vacuum and exchanging by quanta with the zero-point vacuum fluctuations.

\begin{acknowledgements}
The author thanks Denise Puglia, Mike Cavedon, and Pat Noland for useful and valuable remarks and offers.
The author thanks also Miss Pipa (Quantum Portal administrator) for preparing a program drawing Fig.~\ref{fig=1}.
The author thanks the FOOP's reviewers for the constructive critique and proposals
\end{acknowledgements}

\bibliographystyle{spmpsci}      

%
%

\end{document}